\documentclass[final,5p,times,twocolumn]{elsarticle}

\usepackage{amssymb}

\usepackage{amsmath}
\usepackage{amsfonts}

\usepackage{natbib}

\def\d{{\rm d}}

\def\A{{\cal A}}
\def\B{{\cal B}}
\def\C{{\cal C}}

\def\I{{\cal I}}

\def\L{{\cal L}}

\def\P{{\cal P}}
\def\U{{\cal U}}
\def\V{{\cal V}}
\def\R{{\cal R}}
\def\T{{\cal T}}
\def\cc{\text{c}}
\def\d{\text{d}}
\def\i{\text{i}}
\def\n{\text{n}}
\def\s{\text{s}}
\def\x{\text{x}}

\def\lp{l_\text{p}}
\def\tp{t_\text{p}}
\def\min{\text{min}}

\biboptions{square,sort&compress}

\begin{document}

\begin{frontmatter}



\title{Dynamical trapping and relaxation of scalar gravitational fields}


\author[abrd,RAL]{C. H.-T. Wang}
\ead{c.wang@abdn.ac.uk}

\author[abrd]{A. O. Hodson}

\author[edin]{A. St J. Murphy}

\author[abrd]{T. B. Davies}

\author[IST,RAL]{J. T. Mendon\c{c}a}

\author[RAL,stra]{R. Bingham}

\address[abrd]{SUPA Department of Physics, University of Aberdeen, King's
College, Aberdeen AB24 3UE, UK}
\address[RAL]{STFC Rutherford Appleton Laboratory, Chilton, Didcot, Oxfordshire OX11 0QX, UK}
\address[edin]{SUPA School of Physics and Astronomy, University of Edinburgh, Edinburgh, EH9 3JZ, UK}
\address[IST]{GOLP/Centro de F\'isica de Plasmas, Instituto Superior
T\'ecnico, 1049-001 Lisboa, Portugal}
\address[stra]{SUPA Department of Physics, University of Strathclyde, Glasgow G4 0NG, UK}

\begin{abstract}
We present a framework for nonlinearly coupled scalar-tensor theory of gravity to address both inflation and core-collapse supernova problems. The unified approach is based on a novel dynamical trapping and relaxation of scalar gravity in highly energetic regimes. The new model provides a viable alternative mechanism of inflation free from various issues known to affect previous proposals. Furthermore, it could be related to observable violent astronomical events, specifically by releasing a significant amount of additional gravitational energy during core-collapse supernovae. A recent experiment at CERN relevant for testing this new model is briefly outlined.
\end{abstract}


\end{frontmatter}


\section{Introduction and key observations}

Relativistic theories of gravity with a scalar field $\phi$, that actively rescales the metric tensor $g_{ab}$ used by Einstein's general relativity (GR), have been a subject of fundamental significance in theoretical and experimental physics, astronomy and cosmology \cite{Alsing2012}. Such scalar-tensor (ST) theories of gravity arise universally from unified gauge theories and are important for approaches to quantum gravity involving a conformal symmetry \cite{wang2005}. However, in view of the stringent experimental constraints \cite{Bertotti2003} within the solar system on the linearly coupled ST theory due to Brans and Dicke \cite{Brans1961, Dicke1962}, recent works have focused on nonlinear theories under extreme gravity. These include inflation in the early Universe \cite{Steinhardt1990, Clifton2012} and strong field effects in neutron stars, pulsars and core-collapse supernovae (SNe) \cite{Damour1993b, Damour1996, Harada1998, Whinnett2000, Sotani2004, Wang2011, Davies2012}.

As is well known, the observed large scale structure of the Universe, including its flatness and horizon, suggests an inflation shortly following the Big Bang. The original ``old'' proposal \cite{Guth1981} involves the trapping of a hypothetical scalar inflaton field in a false vacuum associated with a prescribed potential. As the Universe supercools this metastable vacuum energy functions as a cosmological constant to drive the exponential expansion of cosmic scale, until the end of the first-order phase transition to the true vacuum state through bubble nucleation. Unfortunately, inflation does not end naturally in this scenario and its nucleation also leads to unacceptably large subsequent fluctuations in density and cosmic microwave background (CMB) \cite{GuthWeinberg1983}.

While a ``graceful exit'' from inflation could be achieved through an alternative ``new'' slow-rolling scalar model \cite{Linde1982} without invoking bubble nucleation, at the expense of a new fine tuning of the inflaton potential, such an option is considered to be {\it ad hoc}. Likewise, other inflation models such as chaotic inflation suffer from various fine tuning drawbacks \cite{LaSteinhardt1989}.

Remarkably, if ST theory is used to evolve the cosmic scale factor $a(t)$ during inflation, then its resulting power-law growth, as opposed to the exponential growth, would find a graceful exit through the first-order phase transition. The first attempt of such an ``extended inflation'' \cite{LaSteinhardt1989} using the linear theory requires a Brans-Dicke coupling outside the observational bounds \cite{Weinberg1989, Green1996}. Fortunately, this flaw can be surmounted by using general ST theories in ``hyperextended inflation'' \cite{Steinhardt1990}, which retains the power-law growth of the scale factor $a(t)$ and the associated graceful exit. Its continuous development can be found in the recent review \cite{Clifton2012} and references therein.

Here we raise a new avenue of inflation using general ST theories, that not only assimilates the strengths of previous models while avoiding their drawbacks but also offers a number of attractive features. We shall assume the scalar gravitational field to play the role of the inflaton field and postulate no further fields violating the strong energy condition. As with new inflation, a potential is required to allow the scalar to roll downhill during inflation. The slowness of rolling and the initial uphill positioning of the scalar need not be fine-tuned. In contrast, it is ``controlled'' through a novel dynamical trapping mechanism as a result of a generic nonlinear coupling between the scalar and matter. The effective vacuum is initially seized in a state with a large cosmological constant and then relaxes to one with a small cosmological constant in a dynamically controlled manner without bubble nucleation. In a sense the model renews new inflation without fine tuning using ST theory. Nonetheless it recovers GR in the low energy domain of the current epoch. In addition, the dynamical trapping mechanism could be reactivated in certain observable energetic astronomical events such as core-collapse SNe.

The new model is constructed from the general Lagrangian for ST theory \cite{Dicke1962, Wagoner1970}:
\begin{eqnarray*}
\frac{c^4}{16\pi G} \int g^{1/2}\R\,\d^3x + \L + L{}
\label{act}
\end{eqnarray*}
in the Einstein frame, where $a,b=0,1,2,3$, $x^0 = ct$, $\R$ is the scalar curvature of the ``Einstein metric'' $g_{ab}$ with signature $(-, +, +, +)$ and $g = -\det(g_{ab})$,
\begin{eqnarray*}
\L = -\frac{c^4}{4\pi G} \int g^{1/2}
\left[\frac12\,g^{ab}\phi_{,a}\phi_{,b} + V(\phi)\right]{\d^3x}
\label{actphi}
\end{eqnarray*}
is the Lagrangian for the scalar gravitational field $\phi$ with a potential
$V(\phi)$, and $L{}$ is the Lagrangian for matter fields coupled to gravity through the ``physical metric'' $\bar{g}_{ab} = \Omega(\phi)^2g_{ab}$. Similarly, in what follows, we shall refer to a barred quantity as a ``physical'' quantity, which is defined with respect to $\bar{g}_{ab}$ as opposed to ${g}_{ab}$. The foregoing conformal mapping uses a coupling function $\Omega(\phi)$, with a coupling strength between the scalar and matter proportional to
$
\alpha(\phi)=\partial{\ln \Omega(\phi)}/\partial{\phi}
$.
The function $\ln \Omega(\phi)$ depends linearly on $\phi$ in the original ST theory by Brans and Dicke \cite{Brans1961,Dicke1962}, but becomes nonlinear in a general ST theory.
Strictly speaking, the gravitational constant $G$ here is defined  with respect to $g_{ab}$. However, anticipating $g_{ab}\to\bar{g}_{ab}$ in the late Universe, we shall simply evaluate $G$ as usual, and use it to define the Planck length $\lp$ and time $\tp$ in the normal way.

The field equations for $g_{ab}$ and $\phi$ follow from varying the total Lagrangian as follows
\begin{eqnarray}
G_{ab} &=& \frac{8\pi G}{c^4}\, (\T_{ab} + T_{ab}{})
\label{geq}
\\
\Box\, \phi &=& V'(\phi) - \frac{4\pi G}{c^4}\, \alpha (\phi)T{}
\label{phieq}
\end{eqnarray}
where $G_{ab}$ is the Einstein tensor and $\Box$ is the Laplace-Beltrami operator with respect to $g_{ab}$, ${\T}^{ab} = 2{{g}}^{-1/2}{\delta \L}/{\delta {g}_{ab}}$ is
the effective Einstein stress tensor for $\phi$,
$
{T}^{ab}{}
=
2{{g}}^{-1/2}{\delta L{}}/{\delta {g}_{ab}}
$
is the Einstein stress tensor for matter,
with the trace ${T}=g_{ab}\,{T}^{ab}{}$. They are related to the physical stress tensor for matter and its trace by
\begin{eqnarray}
{T_{ab}}{}
=
\Omega^2\, \bar{T}_{ab}{},
\hspace{10pt}
{T}=
\Omega^4\, \bar{T}{}
\label{Tm}
\end{eqnarray}
via the conformal mapping between ${g}_{ab}$ and $\bar{g}_{ab}$.

We shall now apply the above framework to cosmology with a view to scalar gravity induced inflation using parameters relevant for energetic astrophysical events in the present epoch. Motivated by the recent interest in the quadratically coupled scalar gravity in neutron stars and core-collapse SNe \cite{Damour1993b, Damour1996, Harada1998, Whinnett2000, Sotani2004, Wang2011, Davies2012}, we shall  consider the second order coupling specified by:
\begin{eqnarray*}
\Omega(\phi) = e^{\frac12\beta\phi^2}
\hspace{8pt}
\text{i.e.}
\hspace{8pt}
\alpha(\phi) = \beta\phi
\label{abeta}
\end{eqnarray*}
for some constant $\beta$.
We adopt the scalar potential
\begin{eqnarray*}
V(\phi) = \xi\phi^\sigma + \frac{\Lambda_0}2
\label{abeta}
\end{eqnarray*}
with positive constants $\xi$, $\sigma$ and $\Lambda_0$. Here $\sigma$ will be a large even integer to accommodate a rolling potential having an asymptotic freedom for small $\phi^2$ in the late Universe as GR is recovered with a cosmological constant $\Lambda_0$ \cite{Wagoner1970}. Current astronomical observation in relation to dark energy places $\Lambda_0 = 10^{-122}\, \lp^{-2}$ \cite{Barrow2011}. Since this value has negligible effects on inflation and core-collapse SNe, we shall neglect $\Lambda_0$ in the present analysis for simplicity.

From Eq. \eqref{phieq} we see that, for $\beta<0$, a large $T$ due to a high mass-energy density tends to induce an instability of $\phi^2$ towards developing a large value. If $T$ does not depend on $\phi$, then this instability may ultimately be limited by $V(\phi)$ through an increased scalar potential, as will be shown to be the case in the following simple model for inflationary cosmology. This results in a trapped $\phi$ in dynamical equilibrium. Furthermore, in a more general inhomogeneous space, the $\phi$ dependence of $T$ can induced an additional stabilization of large $\phi^2$ before the above limiting effect of $V(\phi)$ is activated, as will be demonstrated in another simple model below to illustrate its implications on core-collapse SNe.  The dynamical balance between these two competing forces on the right hand side of \eqref{phieq} therefore leads to the trapping and relaxation of the scalar gravity in high energy environments (e.g. early Universe, core-collapse) and low energy environments (e.g. late Universe, solar system) regimes respectively. Apart from these guiding principles, the precise forms of $\Omega(\phi)$ and $V(\phi)$ are not critical.
A form of trapping scalar fields through an effective potential is first observed in \cite{Khoury2004}, which has been used in a dark energy model in \cite{Brax2004}, with recent developments reviewed by \cite{Clifton2012}. Unlike \cite{Khoury2004}, however, here we consider $V(\phi)$ that increases towards a large $\phi^2$ and a coupling function with a nonlinear $\ln \Omega(\phi)$.


\section{Cosmic inflation driven by trapped scalar gravity}

To focus on the salient effects, we shall neglect the spatial curvature parameter and associate the Einstein metric $g_{ab}$ with the Robertson-Walker spacetime
with the Hubble parameter
$
H(t)={\dot{a}}/{a}
$, the Einstein equation \eqref{geq} yields the Friedmann equations
\begin{eqnarray*}
H^2
&=&
\frac{8\pi G}{3c^2}\, (\U + u)
\label{F1}
\\
\frac{\ddot{a}}{a}
&=&
-\frac{4\pi G}{3c^2}
\left(
\U + 3\P + u + 3p
\right)
\label{F2}
\end{eqnarray*}
with over-dot denoting time derivative.
Here $\U$ and $\P$ are the effective energy density and pressure of $\phi$ derived from $\T^{ab}$ in the standard manner and
$
u = \Omega^4\bar{u}
$
and
$
p{} = \Omega^4\bar{p}{}
$
are the Einstein energy density and pressure of matter, in terms of the physical
energy density $\bar{u}$ and pressure $\bar{p}$ of matter respectively, obtained from Eq. \eqref{Tm}.

As the equation of state for matter, we follow the standard parametrization $\bar{p}{} = w\,\bar{u}{}$ for some constant $w$ so that
\begin{eqnarray*}
\bar{T} = -\bar{u}{}+3\bar{p}{}
=
-(1-3w)\,\bar{u}{}.
\end{eqnarray*}
Since we do not expect matter to drive inflation, we shall be primarily interested in the range $-1/3 < w < 1/3$.
After some algebra, the second Friedmann equation take the form:
\begin{eqnarray*}
\dot{H}
+\frac32(1+w)H^2
+\frac12(1-w)\dot\phi^2
-c^2(1+w)\xi\phi^\sigma
=0.
\label{Eqn1}
\end{eqnarray*}
Under the trapping and quasistatic rolling of $\phi$, we have $\dot\phi\approx0$ and  $\ddot\phi\approx0$, and hence the scalar field equation \eqref{phieq} and the above equation become
\begin{eqnarray*}
3(1-3w) H^2 \beta
-
2c^2(1-3w) \beta\xi\phi^{\sigma}
+
2c^2\xi\sigma\phi^{\sigma-2}
&=& 0
\label{Eqn2a}
\\
\dot{H}
+\frac32(1+w)H^2
-c^2(1+w)\xi\phi^\sigma
&=&0.
\label{Eqn1a}
\end{eqnarray*}
Furthermore, by neglecting $\dot{H}$ in the above equation under the quasistatic assumption, we see that the trapped scalar satisfies
\begin{eqnarray*}
\phi
\approx
\left(\frac{3H^2}{2c^2\xi}\right)^{1/\sigma}.
\label{Eqn2aa}
\end{eqnarray*}

It is convenient to introduce $Q(t)=q(t)+1$ using the decelerating parameter $q(t)$,
so that we can rewrite the above time evolution equation for the Hubble parameter in a neat form:
\begin{eqnarray*}
(H^{-1})\dot{\;}
=
Q(t)
\label{dHoverH1}
\end{eqnarray*}
using the following expression for $Q(t)$:
\begin{eqnarray*}
Q(t)
=
\frac32\,\frac{\sigma(1+w)}{\sigma-\beta(1-3w)\phi(t)^2}.
\label{Q}
\end{eqnarray*}
During inflation we have $q(t) < 0$ and so $Q(t) < 1$, which through the above equation implies an approximate lower bound on $\phi^2$:
\begin{eqnarray*}
\phi_\x^2 = \frac{\sigma (3w + 1)}{2\beta(3w-1)}.
\label{acnd}
\end{eqnarray*}
Once $\phi^2$ reaches $\phi_\x^2$ from above, inflation exits.
The equivalent exit value for the Hubble parameter then follows as
\begin{eqnarray*}
H_\x = c \sqrt{\frac{2\xi}3}\left[\frac{\sigma (3w + 1)}{2\beta(3w-1)}\right]^{\sigma/4}.
\label{Hcnd}
\end{eqnarray*}

There are two important limiting cases: As $w \to 1/3$, describing radiation, $H_\x \to \infty$ implying no inflation, consistent with cosmology where inflation preceding the radiation era. On the other hand, as  $w \to -1/3$, violating the strong energy condition, $H_\x \to 0$ and inflation never ends, as such exotic matter is capable of driving inflation.

Suppose the scalar is initially trapped at time $t_\i$ to be $\phi_\i = \phi(t_\i)$ with the corresponding Hubble parameter $H_\i=H(t_\i) > H_\x$ and  $Q_\i=Q(t_\i)$, before $\phi$ slowly rolls towards zero. Then the time evolution of the Hubble parameter can be described by
\begin{eqnarray*}
\frac1{H(t)}\approx Q_\i(t-t_\i)+\frac1{H_\i}.
\label{approxHinv}
\end{eqnarray*}
This in turn yields the approximate power law expansion for the scale factor:
\begin{eqnarray*}
{a(t)}
\approx
{a_\i}\left[
{Q_\i H_\i (t-t_\i)+1}
\right]^{1/Q_\i }.
\label{asol}
\end{eqnarray*}
Denote the exit time by $t_\x$ so that $H_\x = H(t_\x)$ and $a_\x = a(t_\x)$.
From the above relations we see that over the duration
\begin{eqnarray*}
t_\x - t_\i \approx \frac1{Q_\i}\left(\frac1{H_\x} - \frac1{H_\i}\right)
\label{dtH}
\end{eqnarray*}
the expansion of $a(t)$ has the e-folding of
\begin{eqnarray*}
\ln\frac{a_\x}{a_\i}
\approx
\frac1{Q_\i}\ln
\frac{H_\i}{H_\x}.
\label{efold}
\end{eqnarray*}

As a numerical example, if we take
$
w=0,
\;
\beta = -4,
\;
\sigma=20,
\;
\xi=10^{-22} \,\lp^{-2}
\;
H_\i = 10^{-3} \,\tp^{-1}
$
then we see that the e-folding $\approx 80$ over the period $\approx 10^{10}\, \tp$, which are compatible with the essential requirement for inflation.

\section{Trapping of gravitational scalar in core-collapse supernovae}
In the usual approach to inflation, it seems inconceivable to ask for observational effects, apart from imprints on cosmology such as CMB anisotropies and density perturbations. However, since we are identifying inflaton field with scalar gravitational field, it is imperative to ask whether a related scalar gravitational effect would show up in astronomical events. As discussed above, such events necessarily involve high energy and density. Specifically, we shall address the core-collapse SN problem and assess how the ST theory relevant for inflation may count for (some of) the apparently missing energy required to explain the observed powerful explosions \cite{Janka2012}.

To this end, we consider an additional contribution to gravity due to $\phi$ in the Newtonian limit. We therefore approximate $g_{ab}$ by the Minkowski metric and assume a small scalar gravitational field. Accordingly, we will ignore the potential $V(\phi)\approx 0$ as $\sigma \gg 1$ and treat $\phi$ as massless as per the asymptotic freedom of $V(\phi)$ discussed above. We expect nontrivial effect to come from the coupling function $\Omega(\phi)$ in the presence of high density and adopt $\beta<0$ as in the inflation model above.

To capture the essential dynamical trapping of $\phi$ and its effect on gravitational binding energy, we consider an idealized spherical core-collapse model, in which the gravitational collapse of an iron core of a heavy star is suddenly halted as an incompressible neutron core of radius $R$ with a constant (nuclear) density $\bar\rho$ is formed. From this moment, we shall establish the time evolution of the scalar field, with the associated energy shift and relevant timescale. We regard the collapse to be so rapid, it is not until the neutron core is just formed, that the scalar inside the core starts to evolve from a small cosmological value towards a new equilibrium. The final equilibrium state of the scalar in our analysis below recovers features of ``scalarization'' discussed in the literature \cite{Damour1993b, Damour1996, Harada1998, Whinnett2000, Sotani2004}. We also note that the theoretical time dependent estimates below are consistent with numerical simulation results in a different context reported in \cite{Novak1998, Novak2000}. Our analytical approach to the dynamical trapping of scalar gravitational field offers fresh new insight and simplicity for the present investigation.

Given the high density of the neutron core, we shall neglect density outside the core. Using the ST theory in the Newtonian limit, the total gravitational potential depending on the radial distance $r$ from the core center is derived from $\bar{g}_{00}$ to be the sum
$
\Phi = \Phi_\n + \Phi_\s
$
of the Newtonian gravitational potential
$
\Phi_\n = ({2\pi}/{3})\, G \bar{\rho}(r^2 - 3R^2)
$
and scalar gravitational potential
$
\Phi_\s = (1/2)\, c^2 \beta\phi^2
$.
Using the quasistatic pressure inside the neutron core satisfying
$
\nabla \bar{p} = -\bar{\rho}\nabla\Phi
$
and $\bar{p}=0$ on the surface, we see that
$
\bar{p}(r) = -\bar{\rho}[\Phi(r)-\Phi(R)]
$.
By neglecting traveling scalar waves inside the neutron core, we represent $\phi(r,t)$ for $r < R$ as
\begin{eqnarray*}
\phi(r,t) = \varphi(t)\frac{2R}{\pi r}\sin\frac{\pi r}{2 R}
\label{wphi0}
\end{eqnarray*}
where $\varphi(t) = \phi(0,t)$ denoting the value of the scalar field at the center of the neutron core.
This ansatz arises as the scalar equation \eqref{phieq} can be solved in our present configuration with a constant $T$ using a sinusoidal $r \phi(r,t)$, which vanishes at $r=0$ for a finite $\phi$. Outside the core where $T=0$ the wavelength of the free $r \phi(r,t)$ is assumed to be much larger than the radius of the core and so the $r$ derivative of $r \phi(r,t)$ is negligible on this scale at $r=R$. Without traveling scalar waves inside the core the lowest standing wave mode is therefore represented above.
This is then refined by substituting back into \eqref{phieq} using \eqref{Tm} and $\bar{T}=-c^2\bar{\rho}+3\bar{p}$ with $\bar\rho$ and $\bar{p}$ above, and average this equation over the volume of the neutron core.  The approximate scalar energy inside the core can be calculated using $\T^{00}$ to be
\begin{eqnarray*}
E = \frac12\I\dot\varphi^2 + \V(\varphi)
\label{EV}
\end{eqnarray*}
using the effective potential
\begin{eqnarray*}
\V(\varphi)=\frac12\A\varphi^2+\frac14\B\varphi^4
\label{effV}
\end{eqnarray*}
in terms of the following coefficients:
\begin{eqnarray*}
\I &=&
\frac{\pi}2\,{c^2R^3}/{G}
\nonumber
\\
\A
&=&
\frac{c^4R}{2G}
+
\beta\,({2.6\, c^2 R^3 \bar{\rho}}-7.7\,{G R^5\bar{\rho}^2})
\nonumber
\\
\B
&=&
\beta^2(4.1\,c^2 R^3 \bar{\rho} - 8.9\,G\bar{\rho} R^5).
\end{eqnarray*}
The first and second terms of $\A$ and the term $\B$ originate from the $r$-derivative of $\phi$, density and pressure of the core respectively. Here $\beta$ is related to the second derivative of $\ln\Omega(\phi)$ rather than further details of $\Omega(\phi)$. On the surface of the core, the outgoing wave boundary condition applies. The power carried away by the scalar waves can be calculated using $\T^{0r}$ to be
\begin{eqnarray*}
P
&=&
\frac{4\, c^3 R^2}{\pi^2 G}  \dot\varphi^2.
\label{Tphiflx1}
\end{eqnarray*}
The conservation of scalar energy
$
\dot E+ P=0
$
then yields the equation of motion for the amplitude $\varphi$ as
\begin{eqnarray}
\I\,\ddot\varphi
+\A\varphi
+\B\varphi^3
+\C\dot\varphi
=0
\label{eom}
\end{eqnarray}
where
$
\C = (4/\pi^2)\,{c^3 R^2}/{G}
$.
While $\B > 0$ is generally satisfied, $\A$ can be negative if $\beta$ is below the critical value:
\begin{eqnarray*}
\beta_\cc
=
-\frac{{c^4}}
{{5.2\, c^2  G R^2\bar{\rho}}-15.\,{G^2R^4\bar{\rho}^2 }}.
\label{bc}
\end{eqnarray*}
Then $\V(\varphi)$ becomes Higgs-like having local minima at:
$
\varphi_\min^2=-{\A}/{\B}
\label{vpmin}
$
with the minimum
$
\V(\varphi_\min) = -{\A^2}/{4\B}
\label{Vmin}
$.
Since the relaxation time for Eq. \eqref{eom} from $\varphi\approx0$ is $-\C/\A$ the time $t_0$ for $\varphi(t)$ to evolve from a small cosmological value $\phi_0$ to $\varphi_\min$ can be estimated using the relation
\begin{eqnarray*}
t_0 \approx -\frac{\C}{\A}\ln\frac{\sqrt{-\A/\B}}{\phi_0}.
\label{t0}
\end{eqnarray*}

Let us now take again $\beta=-4$ as with the inflation example above, together with neutron core of mass $2.2 M_\odot$ and $R = 12$ km. Then $\beta_\cc = -3.7 > \beta$ and so scalarization will occur, leading to the dynamical trapping of scalar gravitational field in timescale $t_0$ depending on the cosmological value $\phi_0$. We see that $t_0 \gtrsim 10$ ms for $\phi_0\lesssim10^{-10}$, a value compatible with current experimental tests.

In this case, the trapped scalar has $\phi_\min^2\approx0.01$ at the core center, and through $\V(\varphi_\min)$, we find that this trapping releases $1.9$ foe, or $0.1\%$ of $M_\odot c^2$, of gravitational energy, which is significant for an SN. This model therefore suggests that the discussed dynamical trapping and relaxation of scalar gravitational field could play an important role in explaining both inflation and delayed re-energization in core-collapse SNe.

An opportunity potentially even exists to test the present model, since the amount of $^{44}$Ti ejected from an SN, and thus available for detection by gamma-ray sensitive satellites \cite{Grebenev2012}, is thought to be a property of the underlying explosion mechanism. As the present mechanism would be expected to affect the amount ejected a detection could discriminate between models. A recent experiment conducted at the CERN-ISOLDE facility \cite{Murphy2012} has attempted to remove the major nuclear physics uncertainty, enhancing the chances of such a measurement being made.
The authors are grateful to EPSRC and STFC/CfFP for partial support.
CW and AM acknowledge the hospitality of CERN, where part of this work is carried out.
AH thanks RSE for a Cormack Award.

\end{document}